# High Repetition-Rate Femtosecond Stimulated Raman Spectroscopy with Fast Acquisition

## Matthew N. Ashner and William A. Tisdale*

*Department of Chemical Engineering, Massachusetts Institute of Technology, 77 Massachusetts Avenue, Cambridge, MA 02139, USA*
*\*tisdale@mit.edu*

**Abstract:** Time-resolved femtosecond stimulated Raman spectroscopy (FSRS) is a powerful tool for investigating ultrafast structural and vibrational dynamics in light absorbing systems. However, the technique generally requires exposing a sample to high laser pulse fluences and long acquisition times to achieve adequate signal-to-noise ratios. Here, we describe a time-resolved FSRS instrument built around a Yb ultrafast amplifier operating at 200 kHz, and address some of the unique challenges that arise at high repetition-rates. The setup includes detection with a 9 kHz CMOS camera and an improved dual-chopping scheme to reject scattering artifacts that occur in the 3-pulse configuration. The instrument demonstrates good signal-to-noise performance while simultaneously achieving a 3-6 fold reduction in pulse energy and a 5-10 fold reduction in acquisition time relative to comparable 1 kHz instruments.

## 1. Introduction

Femtosecond stimulated Raman spectroscopy (FSRS) is a powerful technique used to observe changes in the vibrational structure of a system as it evolves after optical excitation [1]. FSRS reveals changes in the sample that are encoded in the vibrational spectrum, such as molecular structure rearrangements [2], coupling of specific molecular vibrations to charge-transfer reactions [3-5], and redistribution of electron density due to photoexcitation [6, 7]. FSRS can also be used to understand vibrational energy relaxation in individual vibrational modes by leveraging resonant enhancement of the FSRS signal [8].

In FSRS, one laser pulse, the actinic pump, optically excites the sample. After a controllable time delay, a pair of pulses consisting of a picosecond, narrow bandwidth Raman pump pulse and a broadband white light probe pulse interrogate the system through stimulated Raman scattering [Fig. 1(a)]. The FSRS spectrum is measured as pump-induced changes in the probe spectrum as the pump beams are modulated [1].

A challenge hindering broad application of FSRS is small signal intensities. Raman scattering is a weakly occurring nonlinear optical phenomenon, and FSRS measures small changes in the already-weak signal induced by the actinic pump. As a result, FSRS experiments often require long acquisition times under exposure to high pulse energies to achieve enough signal-to-noise to resolve features of interest. Like many ultrafast techniques, FSRS was developed and is generally applied using 1 kHz Ti:sapphire amplifiers [1, 9]. Acquisition times and required pulse energies can be reduced by taking advantage of the higher repetition-rate and exceptional stability of modern Yb-based amplifiers. This principle has been applied to surface-enhanced FSRS [10, 11] and FSRS microscopy [12-14], achieving exceptional signal-to-noise in ground-state FSRS spectra.

Here, we present a time-resolved FSRS instrument based on a 200 kHz Yb amplifier that demonstrates superior signal-to-noise performance while using appreciably lower pulse energies and shorter acquisition times than typical FSRS instruments. The key improvement is the implementation of a higher frame-rate detection and faster modulation scheme than 1 kHz systems. We also employ a unique dual-chopping scheme that suppresses scattering artifacts unique to the 3-pulse configuration caused by the higher repetition-rate. We apply this instrument to obtain high-quality time-resolved FSRS data for fluorescein dye in water and demonstrate the improvements in acquisition time made possible by operating at higher pulse repetition-rates. Finally, we discuss new challenges unique to high frequency signal acquisition.

## 2. Experimental details

### 2.1. Optical layout

The optical layout is shown in Fig. 1(c). The light source for the FSRS instrument is a Yb regenerative amplifier (Spectra-Physics Spirit) producing 40 µJ pulses at a center wavelength of 1040 nm, <400 fs FWHM pulse duration, and operating at 200 kHz native repetition-rate. Lower repetition-rates can be accessed using an integrated pulse-picker. The laser output is split with a beam splitter (BS), and one arm is passed through a half wave plate and a polarizing beam splitter to control the split ratio between the probe and Raman pump lines.

The probe is generated by self-phase modulation in a yttrium aluminum garnet (YAG) crystal. 2-3 µW of the fundamental is passed through a 750 nm long pass filter to reject second harmonic generated in WP1, and focused with a 100 mm focal length lens into a 4 mm YAG window. The white light spectrum was optimized by adjusting the power input to the continuum generation using WP1, aperturing the spot with an iris, and translating the YAG window. The beam passes through an iris that isolates the uniform center of the continuum and is recollimated using a 100 mm focal length concave mirror. The spectrum is filtered using a 1040 nm notch dichroic (Semrock) to reject the fundamental, and a 700 nm shortpass filter.

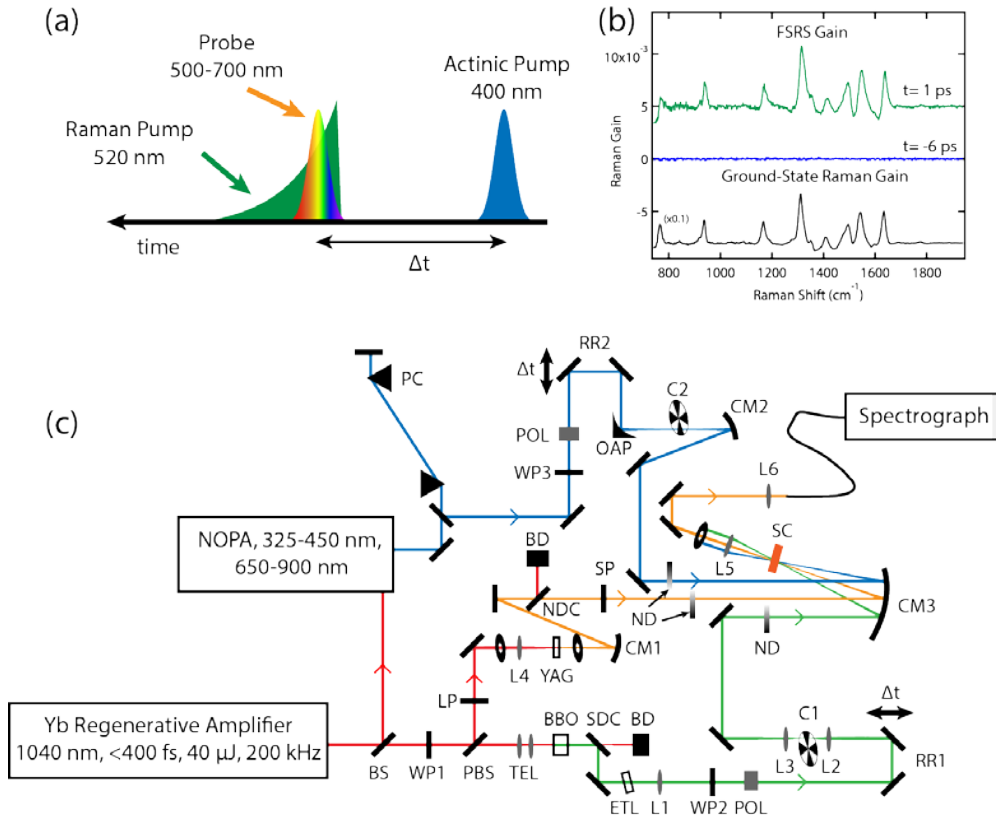

Fig. 1. (a) FSRS pulse sequence. (b) Ground-state FSRS spectrum and time-resolved FSRS gain at -6 ps and +1 ps. (c) Optical layout for the FSRS instrument. BS – beamsplitter; WP1, WP2, and WP3 – half-wave plates; PBS – polarizing beamsplitter; BBO – β-barium borate; YAG – yttrium aluminum garnet; TEL – telescope; SDC – 750 nm shortpass dichroic beamsplitter; ETL – etalon; C1 and C2 – optical choppers; ND – variable ND wheel; L1-L6 – lenses (f = 1000 mm, 100 mm, 50 mm, 100 mm, 200 mm, and 50 mm respectively); POL – polarizer; RR1 and RR2 – retroreflectors; OAP – off-axis parabolic mirror (f = 50.8 mm); CM1, CM2, and CM3 – concave mirror (f = 100 mm, 50 mm, and 200 mm respectively); BD – beam dump; PC – prism compressor; ND – variable neutral density wheel; SC – sample cell.

The portion of the 1040 nm fundamental not used for continuum generation is frequency doubled to 520 nm and spectrally narrowed to produce the Raman pump. First, the beam is reduced in size with a 2-lens telescope and passed through a 4 mm β-barium borate (BBO) crystal to produce the second harmonic at 520 nm. Then, the residual 1040 nm fundamental is removed using a 750 nm shortpass dichroic. The 520 nm light is then passed through a custom designed Fabry-Perot etalon (TecOptics) to produce a pulse with ~3 cm$^{-1}$ FWHM and a temporally asymmetric pulse shape [Fig. 1(a)]. This pulse shape has been shown to reduce the background in resonant stimulated Raman signals, which arises from transient absorption of the Raman pump, and has been shown to slightly enhance the Raman signals [15]. The filtered beam is passed through a 1000 mm focal length lens to compensate for beam divergence over the long path length, followed by a half wave plate and polarizer to control the polarization at the sample. A retroreflector mounted on a motorized stage (Newport) precisely controls the time delay between the Raman pump and probe, which is generally held constant in a given experiment.

The tunable actinic pump is produced by amplification in a commercial non-collinear optical parametric amplifier (Spectra-Physics Spirit-NOPA) pumped from the second arm from BS. For this work, the NOPA was configured with its integrated signal second harmonic generator to output a 50 fs pulse at 400 nm. The output from the NOPA is compressed in a fused silica prism compressor, and passed through a half wave plate (WP3) and α-BBO polarizer. The pump-probe delay is controlled by a retroreflector mounted on a second translation stage (Newport).

All three beams are focused non-collinearly onto the sample using a single large concave mirror (f = 200 mm), and their intensities are controlled by individual variable ND filters. The probe is recollimated by a 200 mm focal length lens and passed through an iris to block the pump beams. A 50 mm lens focuses the probe beam into a fiber, which is coupled to a 500 mm spectrograph equipped with a 1200 gr/mm grating (Princeton Instruments) and a high-speed 1024 pixel linear CMOS camera (Ultrafast Systems).

*2.2. Detection and Electronic Synchronization*

The high repetition-rate system employs a CMOS camera that can be operated at up to 9.7 kHz, allowing for faster pump modulation and better noise suppression than can be achieved with a slower CCD camera. Although the detection is not shot-to-shot, the camera must still be synchronized with the laser pulse repetition-rate to ensure that a consistent number of pulses is collected per frame. Two choppers synchronized with the camera acquisition modulate the Raman and actinic pumps to collect spectra with all pump on/off combinations. Figure 2(a) shows the layout of the electrical components. A data acquisition (DAQ) counter board (National Instruments) receives a TTL signal from the regenerative amplifier and divides it down by a factor of 23 to trigger the CMOS camera. The camera has a software-controlled collection window that is set to 90 µs, or 18 laser shots, so that light is only collected while the pump beams are completely blocked or unblocked by the choppers to maximize pump on/pump off contrast. The chopper controllers are synchronized with another divided signal generated by the counter board.

If a lower repetition-rate is desired, the counter board is used to generate a trigger signal (synchronized with the camera trigger) to control the laser's integrated pulse picker. Like the initial synchronization to the laser, this ensures that each camera frame acquires the same number of pulses. This strategy also ensures that the laser only emits pulses during the detection window, eliminating unnecessary exposure of the sample to intense laser light.

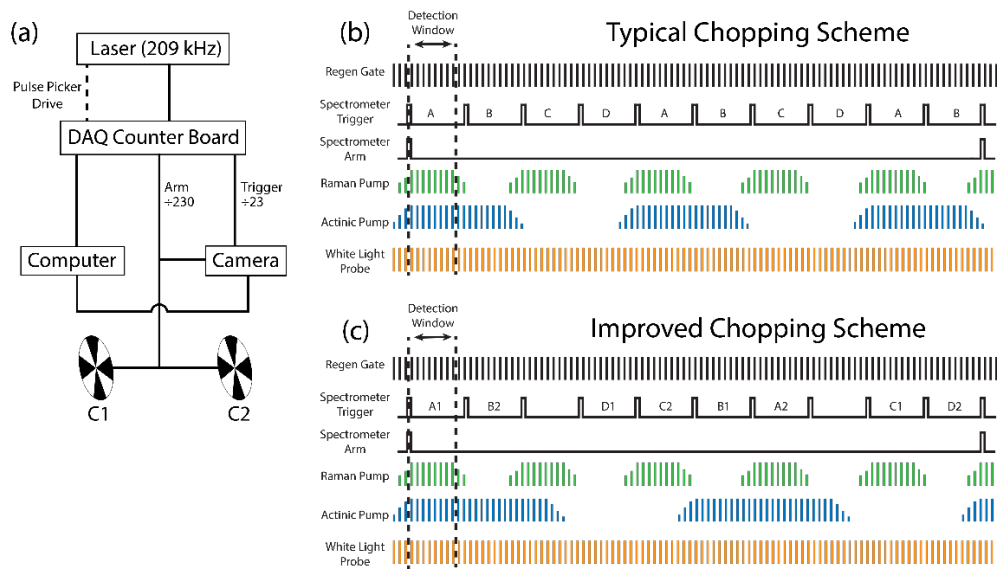

Fig. 2. (a) Schematic layout of the electronics connections. (b) Illustration of a typical data collection scheme applied to a high repetition-rate system depicting the synchronization signals and the modulation of the pump beams. The 4 interleaved spectra are labeled A-D. (c) Illustration of the improved chopping scheme proposed in this work. Again, the 4 interleaved spectra are labeled A-D and their position at the beginning or end of the long chopping phase is denoted 1-2.

The typical chopping scheme has the Raman pump chopped at half the camera frame rate and the actinic chopped at one-fourth of the frame rate so that the probe spectrum for all four pump combinations is collected every four frames [Fig. 2(b)]. However, we found that, at high repetition-rates, artifacts can arise due to differences between the first and second halves of the lower frequency chop (see Section 4.2). To counter this, we introduced an improved chopping scheme, illustrated in Fig. 2(c). In this scheme, a complete cycle is 10 camera frames, with the Raman pump chopped at half of the camera frame rate, and the actinic pump chopped at 1/5 of the frame rate. In each cycle, all 4 pump combinations (labeled A-D) occurring during the beginning and end of the actinic pump chopping phase (labeled 1-2) are acquired, and 2 frames are dropped, for a total of 10 frames.

## 2.3. Sample Preparation

To characterize the performance of our instrument, we used fluorescein dye (Sigma-Aldrich) in water as a sample. A 100 mL reservoir of sample solution was prepared by dissolving the fluorescein in water at a concentration of 6 mM. The pH of the solution was adjusted to 9 using commercial ammonium hydroxide solution. For the measurement, the solution was flowed through a 1 mm path length quartz flow cell (Starna Cells 583.65-Q-1) using a peristaltic pump (Cole-Parmer).

## 2.4. Data Processing

The electronics are run by a custom LabVIEW script that reads the individual frames from the camera system. The software then separates the 10 interleaved spectra in each batch of frames. The paired frames corresponding to the same pulse configurations [both pumps on, actinic pump only, Raman pump only, and both pumps off, A-D in Fig. 2(c)] are averaged to give the 4 probe spectra, and the 2 remaining frames are discarded (e.g. the A1 and A2 frames are averaged together to give the *both-pumps-on* spectrum, A). The ground-state Raman gain ($R_{GS}$) is calculated by dividing the Raman pump on and off spectra with the actinic pump off [Fig.

1(b), black trace],

$$R_{GS} = \frac{C}{D}. \quad (1)$$

For simplicity, and to mitigate the effects of background transient absorption signals[16], the transient FSRS signal is calculated by dividing the Raman gain with the actinic pump on ($R_{ES}$) by the ground-state Raman gain [Fig. 1(b), blue and green traces],

$$FSRS = \frac{R_{ES}}{R_{GS}} = \frac{A/B}{C/D}. \quad (2)$$

The transient FSRS traces contain a sloping background, which is fit using an outlier-rejecting local regression algorithm with a window of 200 points, and subtracted from the spectra. Because the white light probe is not compressed, the resulting traces must be corrected for probe chirp. This is done by collecting a cross-correlation using the optical kerr effect flowing pure water through the flow cell to measure the chirp profile, and applying a correction to the background-subtracted data. The Raman shift axis was calibrated using a standard (toluene).

## 3. Time-resolved femtosecond stimulated Raman spectroscopy

Fig. 3 shows a full time-resolved FSRS trace and selected spectra of fluorescein to demonstrate the performance of the instrument using the improved chopping scheme. Typical FSRS experiments on organic systems employ a Raman pump excitations of 4-10 mJ/cm$^2$ and acquire for 20-60 seconds per time point [2, 3, 6, 8]. The data in Fig. 3 were acquired using a 0.95 mJ/cm$^2$ Raman pump and integrating for 8 seconds per time point. The actinic pump energy density in this experiment was 1.75 mJ/cm$^2$.

If higher pulse energies can be tolerated, then superior signal-to-noise performance is possible. Fig. 4 shows an example of data collected with 4 mJ/cm$^2$ Raman pump and 3.5 mJ/cm$^2$ actinic pump. These data were collected at 100 kHz for 8 seconds per time point, as bleaching caused a marked decrease in signal at 200 kHz – even with the flow cell. The spectra illustrate exceptional signal to noise performance and the ability to resolve weaker Raman features.

## 4. Optimizing signal-to-noise ratio at high repetition-rates

### *4.1. Sample degradation and refresh*

Although the pulse energies in this experiment are lower than in typical FSRS experiments, samples are exposed to significantly higher average powers. In this case, the actinic and Raman pump powers were 4.7 mW and 10 mW, respectively. At these light levels, photobleaching of even robust dyes like fluorescein can be seen in the measurements without appropriate precautions. Figure. 5(a) demonstrates the importance of using a flow cell for solution samples. The flow cell refreshed the illuminated sample volume every 2.8 ms, which was sufficient to mitigate photobleaching effects, but not to refresh the sample between laser shots. If the sample is more sensitive, and photobleaching in a few laser shots is a concern, a higher velocity flow cell, jet, or sample translation should be considered.

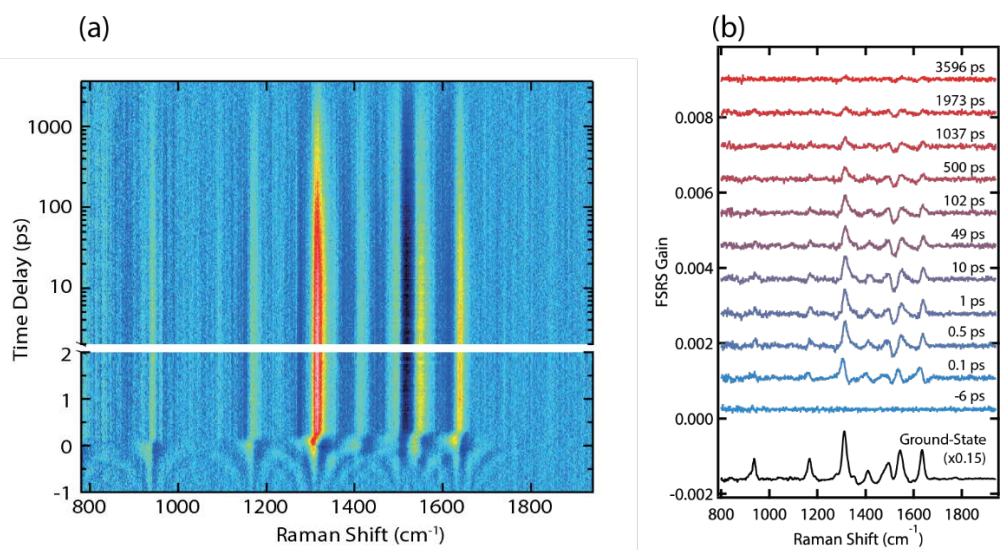

Fig. 3. Low-fluence FSRS data (0.95 mJ/cm$^2$ RP, 1.75 mJ/cm$^2$ AP). (a) Full time-resolved FSRS gain spectrum of fluorescein. (b) Offset FSRS gain spectra at selected time points. The ground state Raman spectrum acquired during the measurement is shown in black at the bottom of the panel, for reference.

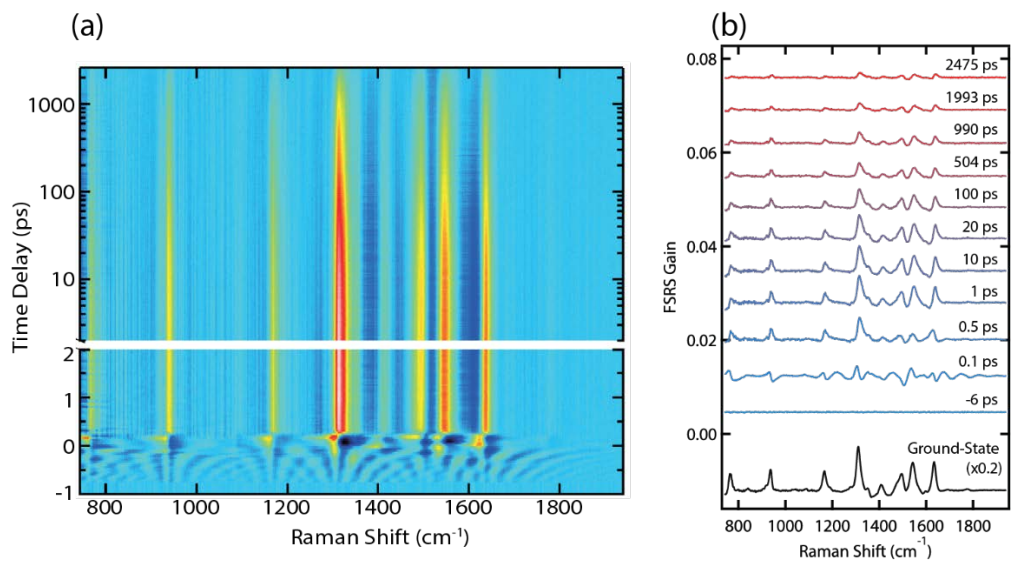

Fig. 4. High-fluence FSRS data (4.0 mJ/cm$^2$ RP, 3.5 mJ/cm$^2$ AP). (a) Full time-resolved FSRS gain spectrum of fluorescein. (b) Offset FSRS gain spectra at selected time points. The ground state Raman spectrum acquired during the measurement is shown in black at the bottom of the panel, for reference.

*4.2. Scatter artifact and improved chopping scheme*

We found that the noise profile when using the typical chopping scheme [Fig. 5(a), orange and green traces] was not random, but rather a systematic artifact that was reproducible under the same experimental conditions. Pixel-to-pixel variation in camera sensitivity is a common source of such artifacts. To correct for this, Challa *et al.* proposed a scanning multichannel detection technique, where the spectrograph grating is moved with each acquisition to shift the spectrum across the camera. The collected spectra are then corrected for the shift and averaged together to smooth out pixel-to-pixel variations [17]. Applying this technique to our system failed to improve the noise floor because the noise pattern shifted together with the Raman spectrum. This indicates that the noise patterns shown in Fig. 5(a) are real high-frequency scattering signals occurring in the experiment.

The source of the noise artifact was determined by examining pairs of spectra acquired under a variety of chopping schemes. Interestingly, the artifact was not associated with any particular pulse combination; instead, it correlated with position in the chopping sequence. To illustrate this, transient absorption spectra were collected using the chopping scheme shown in Fig. 2(b) with the Raman pump blocked. The spectra, shown in Fig. 5(b), reveal that there is a difference between the transient absorption signals in Window A and Window B – i.e. the early and late part of the long chopping phase. This is true in the full FSRS experiment, as well, and is also observed if the chopping frequencies of the actinic and Raman pump beams are swapped.

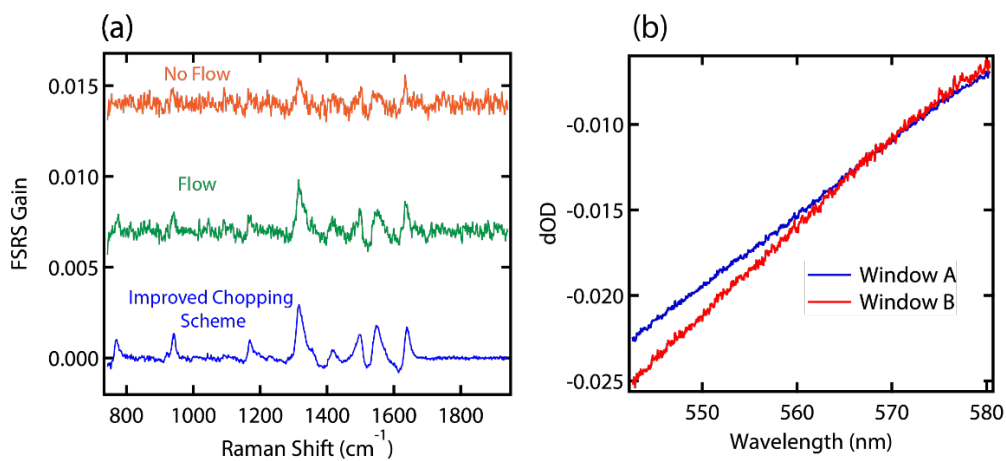

Fig. 5. (a) FSRS gain spectra at t = 6.7 ps acquired with the typical chopping scheme and the flow cell peristaltic pump turned off (orange) and on (green). The blue trace was acquired with the flow cell running and the improved chopping scheme, showing superior signal-to-noise performance. (b) Comparison of transient absorption traces collected in the chopping *window A* (blue) and *window B* (red) phases of the chopping scheme shown in Fig. 2(b) (Raman pump blocked) as described in the main text.

We suspect that the artifact arises from heating in the sample during the actinic pump's long on-phase. Because the laser repetition-rate is faster than the timescale for complete thermal dissipation within the aqueous solution, there is a slow buildup of heat on the 100 µs timescale that leads to corresponding spatial variations in the refractive index and subsequent differences in scattering profiles on the timescale of the chopping frequency. To correct for this effect, we employed the improved chopping scheme illustrated in Fig. 2(c), which samples all combinations of pulses and chopping windows. For example, instead of the *both-pumps-on* combination always occurring at the beginning of the actinic pump's on-phase, it alternates between the beginning and end, allowing for the differences in background scattering to be averaged out. The result is a dramatically improved signal-to-noise ratio even though 20% less light is collected, as shown in the blue trace of Fig. 5(a). The rms noise in the featureless 1700-1850 cm$^{-1}$ region is 5.3x10$^{-5}$, which approaches but it somewhat higher than the shot noise limit of 1.0x10$^{-5}$ for this measurement. The noise level is limited by other, possibly systematic noise sources. For example, modal interference in the optical fiber adds a small systematic noise component [18].

*4.3. Signal-to-noise improvements at high repetition-rates*

Operating at 200 kHz allows for faster data acquisition than is possible with a 1 kHz system. Fig. 6(a) shows a series of spectra acquired for 8 seconds at varying repetition-rates acquired using the low-fluence configuration (0.95 mJ/cm$^2$ Raman pump, 1.75 mJ/cm$^2$ actinic pump). The noise floor markedly improves at higher repetition-rates, as expected. To show that this is primarily a function of increased light throughput, spectra were again acquired at different repetition-rates, but with the acquisition times adjusted such that all measurements had the same total light exposure [Fig. 6(b)]. At 42 kHz and above, the noise floor is constant. At lower repetition-rates, the noise begins to increase. This is likely due to read noise from the CMOS camera, which is not optimized for lower light levels.

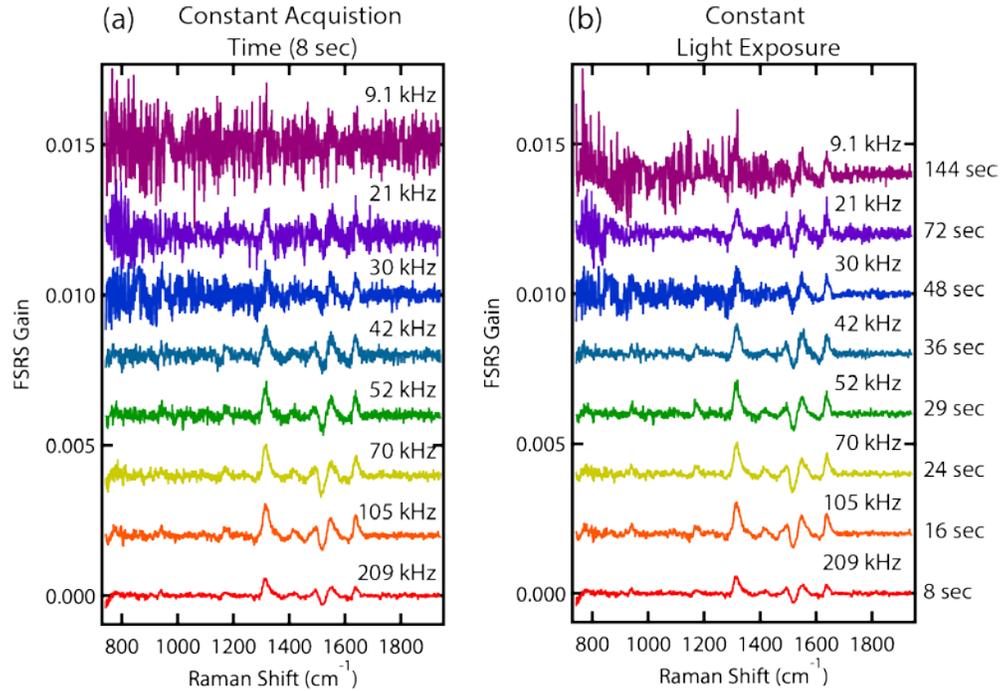

Fig. 6. FSRS gain spectra at t = 6.7 ps as a function of laser repetition-rate, using 0.95 mJ/cm$^2$ Raman pump and 1.75 mJ/cm$^2$ actinic pump pulse fluence. (a) Data collected with constant 8 second acquisition time. (b) Data collected with the integration time scaled so that all acquisitions include the same number of pulses. The integration time was set according to the number of pulses that fall within a camera detection window, and don't scale strictly with the repetition-rate.

Reducing the repetition-rate will also reduce the heating artifact as the sample is exposed to less averaged power, although this is not evident in Fig. 6 because the artifact is corrected by the improved chopping scheme. Comparing to Fig. 5(a), the noise level of the 42 kHz measurement in Fig. 6(a) approximately matches that of the uncorrected measurement at 200 kHz. As such, the heating artifact will stop dominating the noise below a repetition-rate in the high 10's of kHz, although the exact number will depend on the pulse energy, acquisition time, and sample. Although the improved chopping scheme robustly removes the heating artifact from the data, an experimenter operating at high repetition-rates may still find the heating to be unacceptable. The trade-off between signal-to-noise and heating should be considered carefully in each individual experiment, and operating at intermediate repetition-rates may be more appropriate in some cases.

## 5. Conclusion

This work presented the design and construction of a time-resolved femtosecond stimulated Raman spectroscopy system built around a high repetition-rate Yb amplifier and discussed benefits and pitfalls that can be expected from such a system. The system included detection with a high-speed CMOS camera, fast modulation of the pump beams, and an improved chopping scheme that removes probe scatter artifacts in the sample. The instrument achieves performance comparable to other reports in the literature with a 3-6 fold reduction in acquisition time and a 5-10 fold reduction in pulse energy. Faster acquisition time enables experiments on less robust samples and finer resolution of fast dynamics. The experimental improvements implemented can also be applied to other 3-pulse ultrafast techniques implemented at high repetition-rates such as time-resolved sum frequency generation[19], pump-dump-probe [20-22], or pump-push-probe experiments [21-23].

## Funding

This work was supported by the Division of Chemistry and Division of Materials Research, U.S. National Science Foundation, under Award 1452857. M.N.A. acknowledges partial support from the National Defense Science & Engineering Graduate Fellowship Program.

## Acknowledgements

We would like to thank Alexander Morrison, Alexander Sobolev, Alexey Gusev, and Donny Magana from Ultrafast Systems for providing critical information about the cameras and electronics.